\documentclass[aps,prl,twocolumn,showpacs,nofootinbib,preprintnumbers]{revtex4}

\usepackage{color,amsmath,amssymb,subfigure}
\usepackage[dvips]{graphicx}

\begin{document}

\voffset 1.25cm

\title{Cosmic ray spectral hardening due to dispersion in the source
injection spectra}

\author{Qiang Yuan$^{1,2}$, Bing Zhang$^2$ and 
Xiao-Jun Bi$^{1}$}
\affiliation{
$^{1}$Key Laboratory of Particle Astrophysics, Institute of High Energy
Physics, Chinese Academy of Sciences, Beijing 100049, P. R. China \\
$^{2}$Department of Physics and Astronomy, University of Nevada Las Vegas,
Las Vegas, NV 89154, USA
}

\date{\today}

\begin{abstract}

Recent cosmic ray (CR) experiments discovered that the CR spectra
experience a remarkable hardening for rigidity above several hundred GV.
We propose that this is caused by the superposition of the CR energy
spectra of many sources that have a dispersion in the injection
spectral indices. Adopting similar parameters as those of supernova
remnants derived from the Fermi $\gamma$-ray observations, we can
reproduce the observational CR spectra of different species well.
This may be interpreted as evidence to support the supernova remnant
origin of CRs below the knee. We further propose that the same mechanism
may explain the ``ankle'' of the ultra high energy CR spectrum.

\end{abstract}

\preprint{arXiv:1104.3357}

\maketitle

{\it Introduction}---Nearly 100 years after the discovery of cosmic rays
by V. Hess in 1912, several basic questions on CRs, such as the
source(s) and the propagation effects, are still not well understood.
Precise measurements of the CR spectra and observations of
high energy $\gamma$-rays and neutrinos, are of great
importance to approach the answers to these fundamental questions.

There are several major progresses in the CR measurements in recent years.
The balloon-borne experiment Cosmic Ray Energetics And Mass (CREAM)
measured the energy spectra of the major species from proton to iron
in the energy range from tens of GeV/nucleon to tens
of TeV/nucleon with relatively high precision \cite{2009ApJ...707..593A,
2010ApJ...714L..89A}. A remarkable hardening at $\sim 200$ GeV/nucleon
of the spectra of all species was discovered \cite{2010ApJ...714L..89A}.
Most recently the satellite experiment Payload for Antimatter Matter
Exploration and Light-nuclei Astrophysics (PAMELA) reported the precise
measurement about the proton and Helium spectra with rigidity from GV to
1.2 TV \cite{2011Sci...332...69A}. PAMELA data show clearly that the
proton and Helium spectra deviate from the single power-law function above
$\sim 30$ GV with a hardening at rigidity $\sim 200$ GV, which is
basically consistent with the results of CREAM and the previous Advanced
Thin Ionization Calorimeter (ATIC-2) \cite{2007BRASP..71..494P}.
The hardening of the CR spectra challenges the traditional CR acceleration
and propagation paradigm. Models possibly to explain such a spectral
hardening include the multi-component sources \cite{2006A&A...458....1Z},
or the nonlinear particle acceleration scenarios where the feedback of
CRs on the shock is essential (e.g., \cite{1999ApJ...526..385B,
2001RPPh...64..429M,2010ApJ...718...31P}).

{\it Model}---In this work we propose that the hardening of the
observed CR spectra is due to dispersion of the injection
spectra of the CR sources such as supernova remnants (SNRs).
A superposition of the spectra of many sources
with a distribution of the injection spectra would lead to an asymptotic
hardening of the final spectra of CRs \cite{1972ApJ...174..253B}. For
example for power-law injection spectrum $E^{-\gamma}$, if there is a
uniform distribution of the $\gamma$ index $p(\gamma)=\frac{1}{\gamma_2
-\gamma_1}$, the total spectrum will be $\int_{\gamma_1}^{\gamma_2}
E^{-\gamma}p(\gamma){\rm d}\gamma \propto (E^{-\gamma_1}-E^{-\gamma_2})/
\ln E$, which will asymptotically approach the hardest injection spectrum.
Such an effect was adopted to explain the ``GeV excess'' of Galactic diffuse
$\gamma$-rays observed by EGRET \cite{1998ApJ...507..327P,2001A&A...377.1056B}.

Observations of X-ray, GeV and TeV $\gamma$-rays indicate that the SNRs
can accelerate particles (electrons and/or nuclei) up to very high energies.
Those particles are thought to be the most probable sources of the
Galactic CRs. Studies of radio emission spectra of SNRs suggest that
the spectral indices of accelerated particles have a significant
dispersion instead of a uniform value \cite{2001AIPC..558...59G}.

The Fermi satellite observed several SNRs in the GeV band with high
precision \cite{2009ApJ...706L...1A,2010Sci...327.1103A,2010ApJ...712..459A,
2010ApJ...710L..92A,2010ApJ...718..348A,2010ApJ...722.1303A,
2011ApJ...734...28A}. The $\gamma$-ray energy spectra and the coincidence
with molecular clouds of several SNRs suggest that the $\gamma$-rays are
most likely of a hadronic origin, although the leptonic origin is not
ruled out. A broken power-law injection of protons with the break energy
several to tens of GeV seems to describe the $\gamma$-ray data well
\cite{2011MNRAS.410.1577O,2010arXiv1010.1901Y}. The fit to $\gamma$-rays
also indicates that there is a large dispersion of the accelerated CR
spectra in SNRs. Assuming the hadronic origin of the $\gamma$-ray
emission of SNRs, the modeling of the GeV-TeV $\gamma$-ray emission
from eight sources detected by Fermi and Cherenkov telescopes gives
the source particle spectra $\gamma_1\approx 2.15\pm0.33$ and
$\gamma_2\approx 2.54\pm0.44$, for energies below and above the break
respectively \cite{2010arXiv1010.1901Y}.

Here we calculate the superposition effect of CR spectra, assuming that
CRs are originated from SNR-like sources. The injection spectrum of each
source is assumed to be a broken power-law function of rigidity with the
break from several to tens of GV. The spectral indices are assumed to be
Gaussian distributed around some average values. The normalization of
each source is derived assuming a constant total energy of CRs above
1 GeV for all sources. Since the particle spectra inferred from the
$\gamma$-rays might be different from those leaking into the interstellar
space \cite{2010APh....33..160C}, the injection parameters are adopted
through the fit to the CR data instead of the ones inferred from
$\gamma$-ray observations.


We employ the GALPROP code \cite{1998ApJ...509..212S} to calculate the
propagation of CRs, in the diffusive reacceleration frame.
The main propagation parameters are $D_0=5.8\times10^{28}$ cm$^2$
s$^{-1}$, $\delta=0.33$, $v_A=32$ km s$^{-1}$ and $z_h=4$ kpc.
For each major chemical species we use the superposed spectra
of many sources as input and calculate its propagation. The B/C ratio
is found well consistent with data and insensitive to the source
spectrum.

\begin{figure*}[!htb]
\centering
\includegraphics[width=0.65\columnwidth]{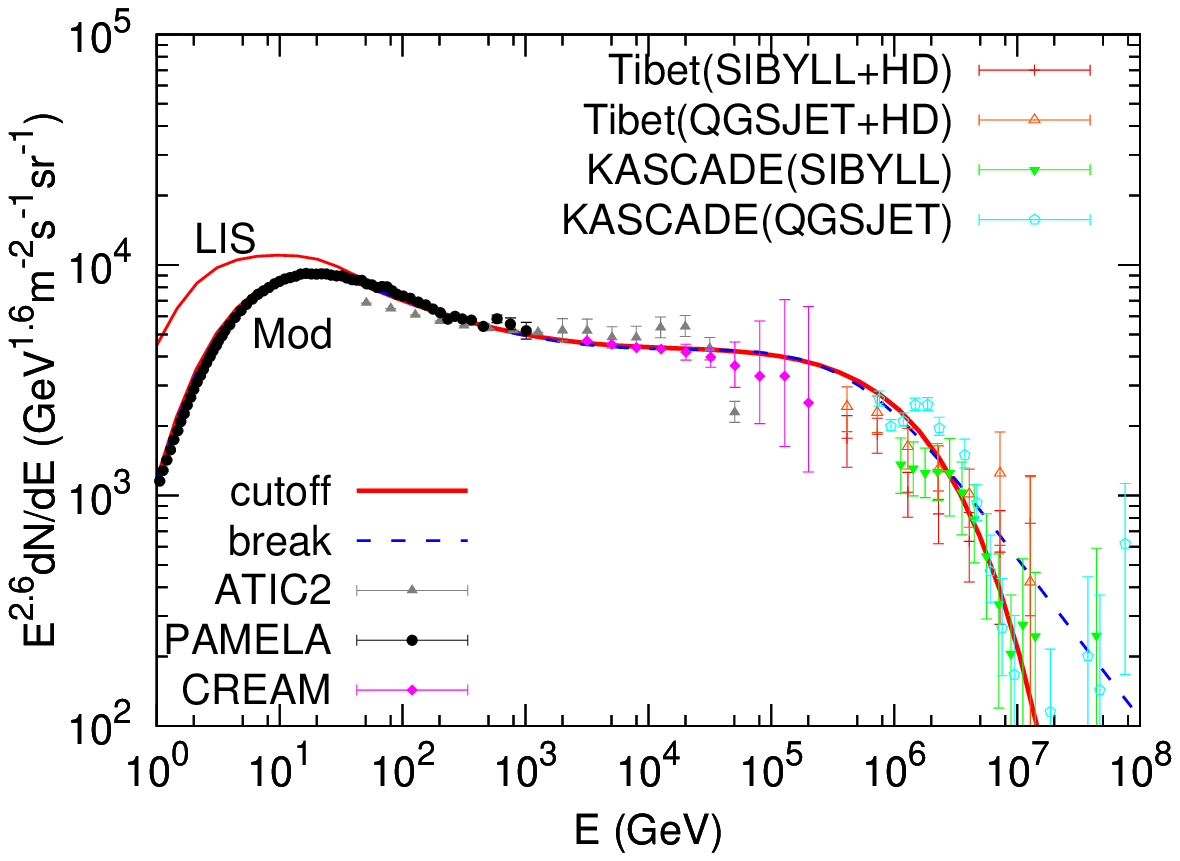}
\includegraphics[width=0.65\columnwidth]{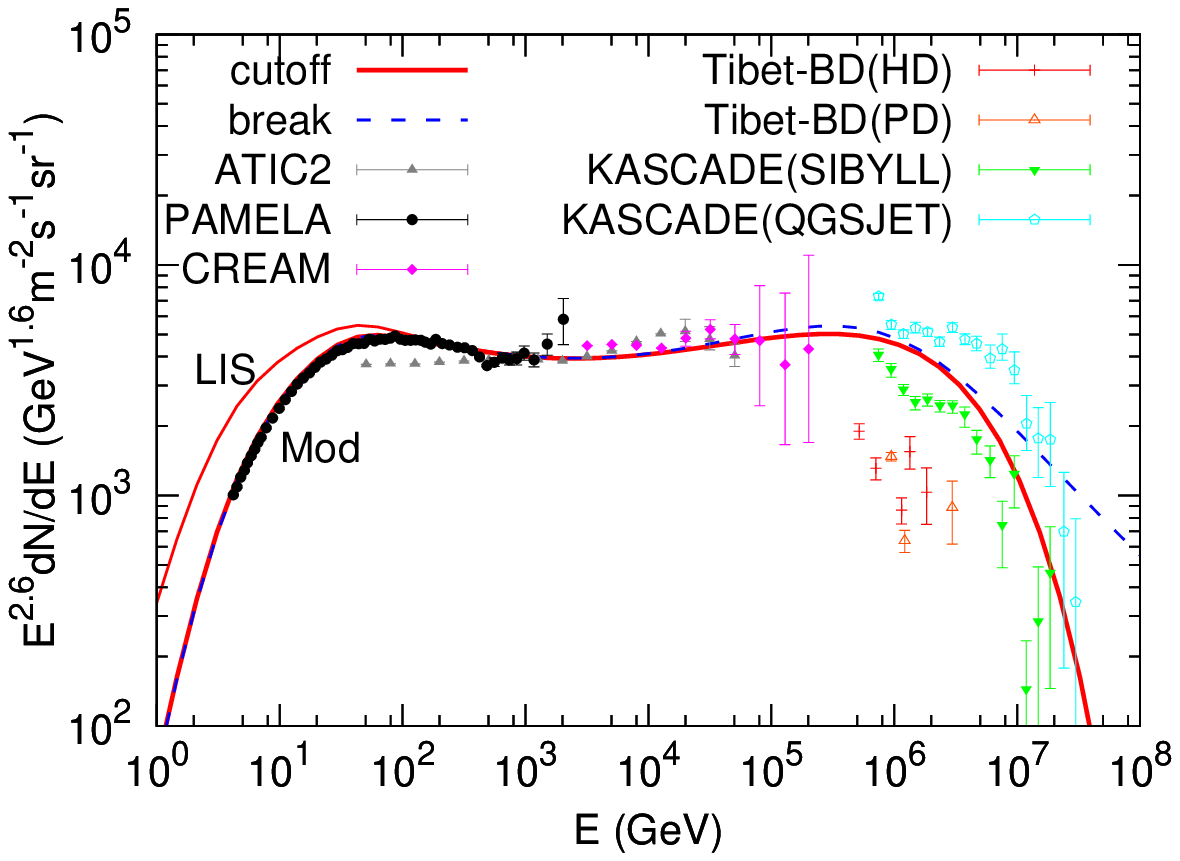}
\includegraphics[width=0.65\columnwidth]{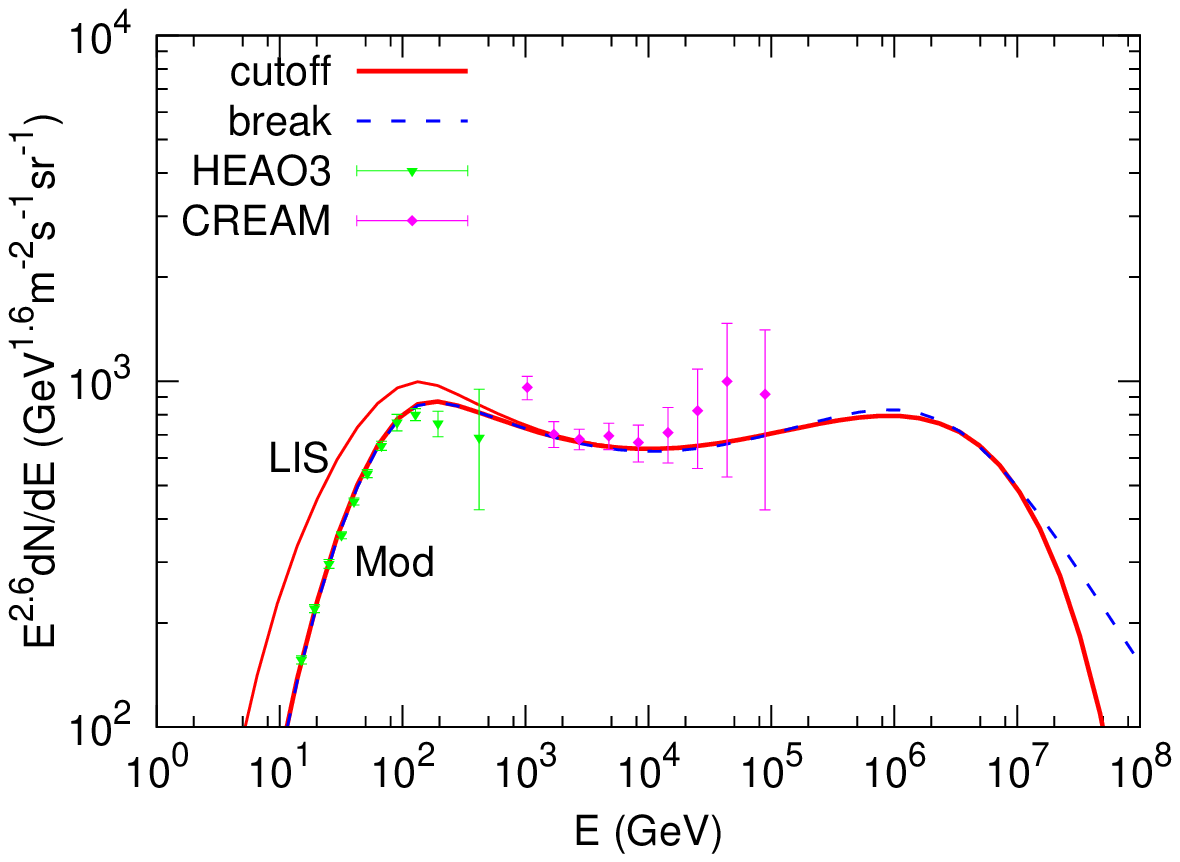}
\includegraphics[width=0.65\columnwidth]{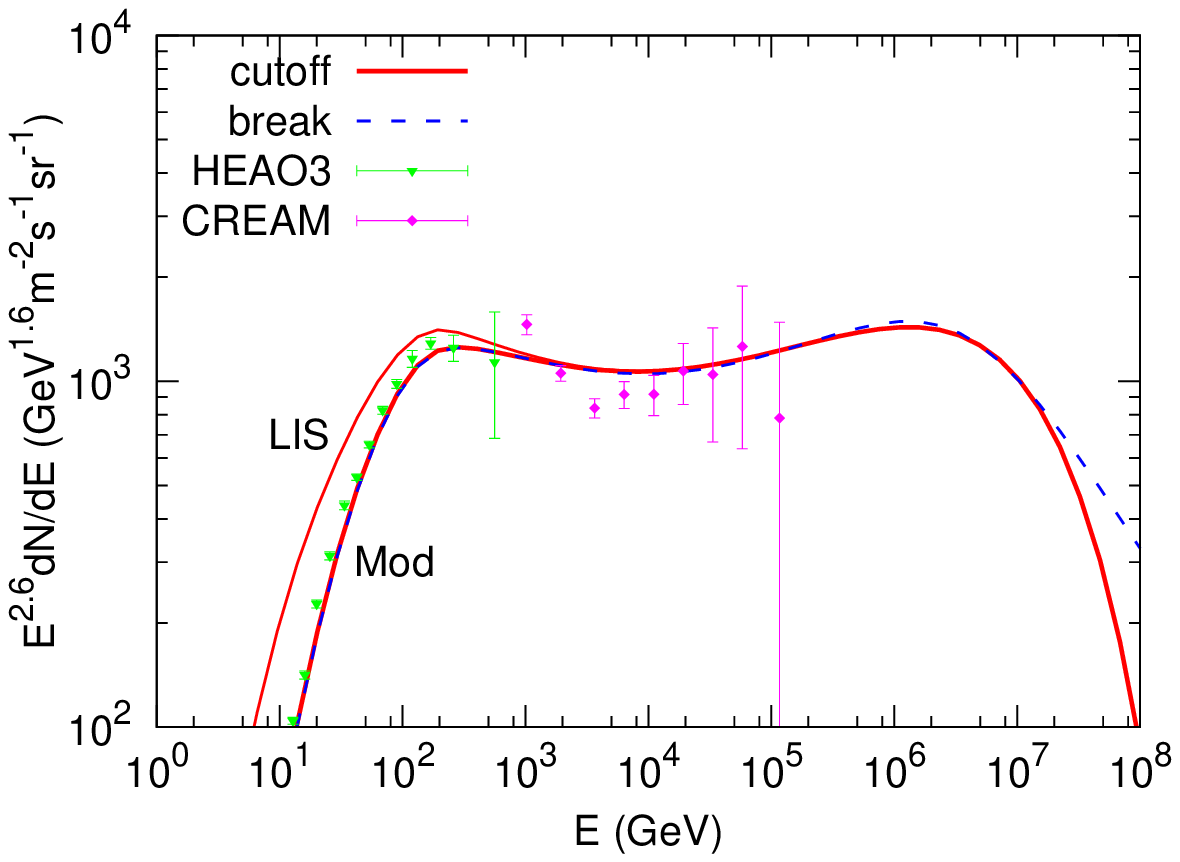}
\includegraphics[width=0.65\columnwidth]{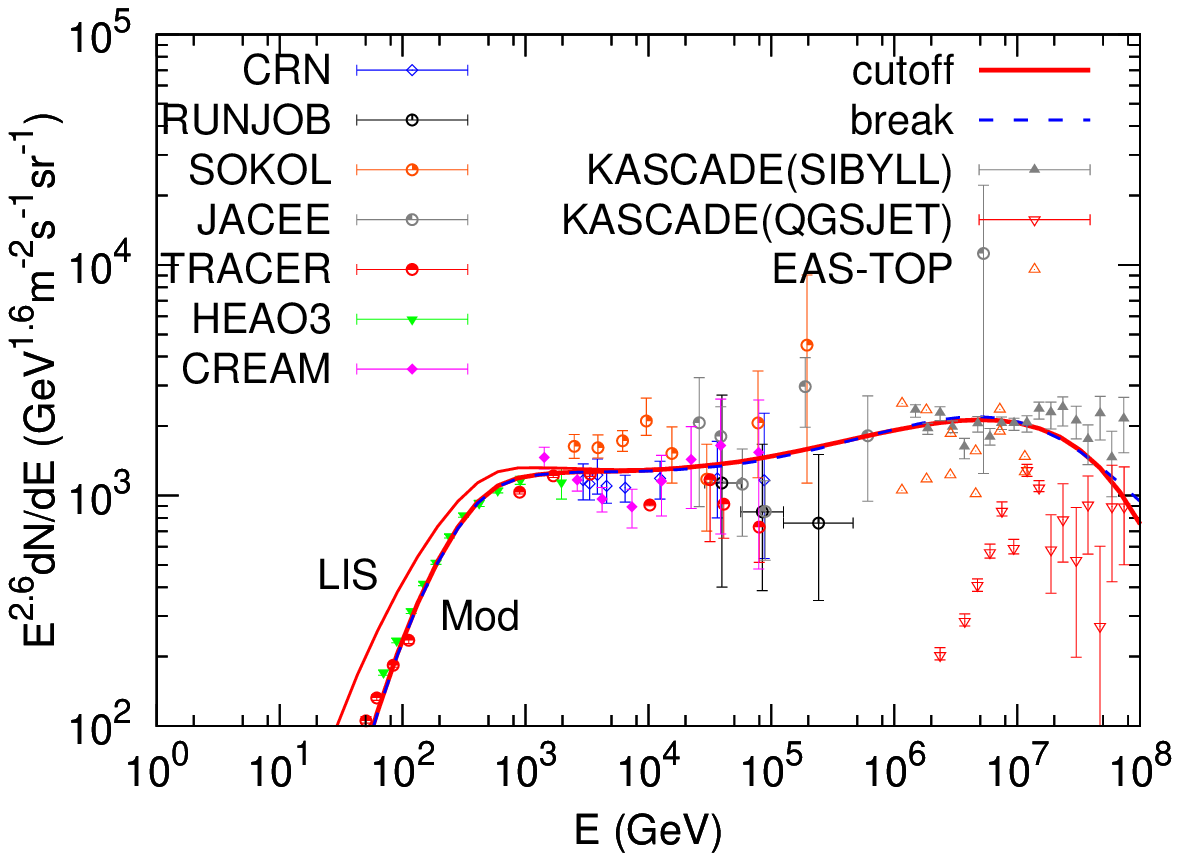}
\includegraphics[width=0.65\columnwidth]{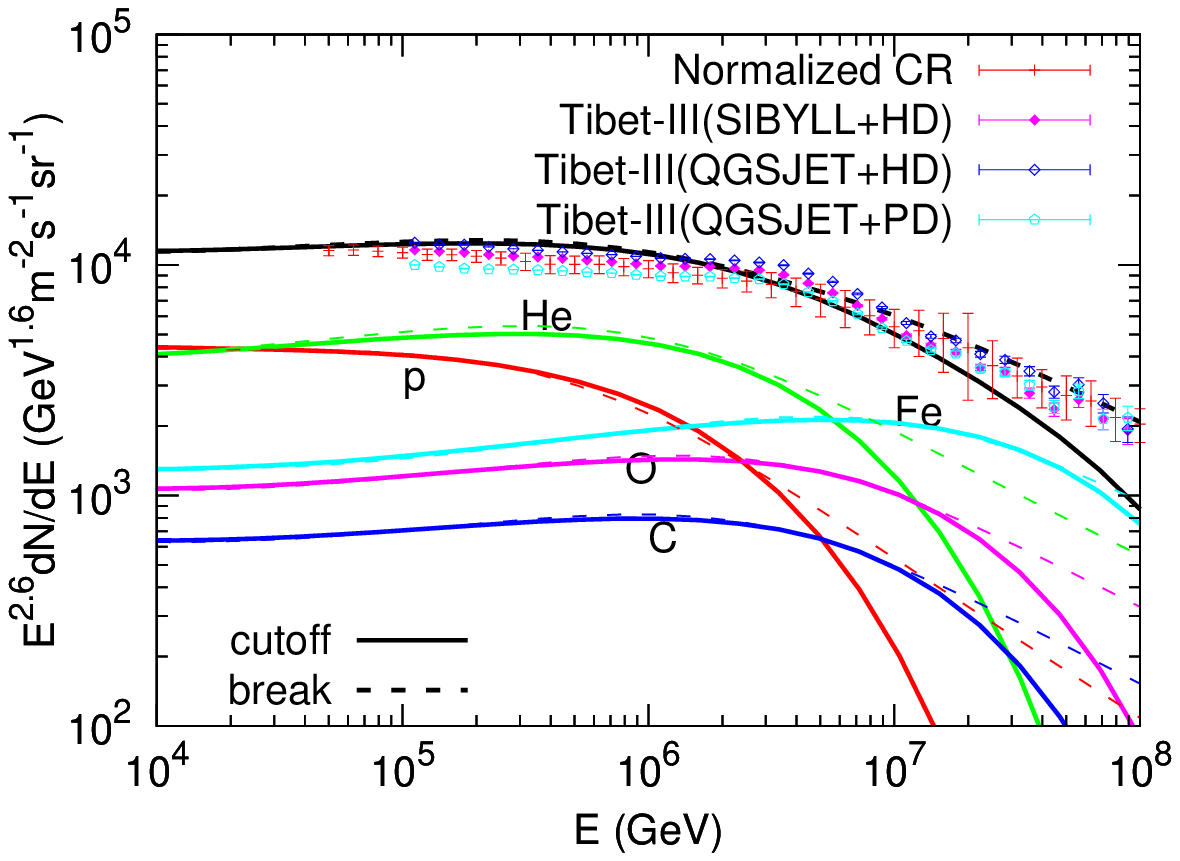}
\caption{Energy spectra of proton (top-left), Helium (top-middle),
Carbon (top-right), Oxygen (bottom-left), Iron (bottom-middle)
and the all-particle one (bottom-right). The local interstellar spectra
are labelled with ``LIS'' and the observed spectra after solar
modulation are labelled with ``Mod''. The solid line in each panel
represents a sub-exponential cutoff behavior of the high energy spectra
around the knee region, while the dashed line is for broken power-law type.
References of the data are---proton: ATIC-2 \cite{2007BRASP..71..494P},
PAMELA \cite{2011Sci...332...69A}, CREAM \cite{2010ApJ...714L..89A},
Tibet \cite{2006JPhCS..47...51A}, KASCADE \cite{2005APh....24....1A};
Helium: ATIC-2 \cite{2007BRASP..71..494P}, PAMELA \cite{2011Sci...332...69A},
CREAM \cite{2010ApJ...714L..89A}, Tibet-BD \cite{2000PhRvD..62k2002A},
KASCADE \cite{2005APh....24....1A}; Carbon: HEAO3
\cite{1990A&A...233...96E}, CREAM \cite{2009ApJ...707..593A};
Oxygen: HEAO3 \cite{1990A&A...233...96E}, CREAM \cite{2009ApJ...707..593A};
Iron: CRN \cite{1991ApJ...374..356M}, RUNJOB \cite{2005ApJ...628L..41D},
SOKOL \cite{1993ICRC....2...17I}, JACEE \cite{1995ICRC....2..707A},
TRACER \cite{2008ApJ...678..262A}, HEAO3 \cite{1990A&A...233...96E},
CREAM \cite{2009ApJ...707..593A}; all-particle: Tibet-III
\cite{2008ApJ...678.1165A}. The normalized all-particle data are derived
by combining all data with a rescale based on the extrapolation of the
direct measurements \cite{2003APh....19..193H}.
}
\label{fig:pamela}
\end{figure*}

For rigidity below $\sim 30$ GV, solar modulation needs to
be considered. The force-field approximation is adopted to model
the effect \cite{1968ApJ...154.1011G}. The modulation potential
$\Phi$ depends on the solar activity, which varies from $\sim 200$ MV
at solar minimum to $\sim 1400$ MV at solar maximum. For the period
when PAMELA operates, the modulation potential was estimated to be $450-550$
MV \cite{2011Sci...332...69A}. According to the fit to the low energy
observational data (see Fig. \ref{fig:pamela}), we adopt $\Phi=550$
MV for proton and Helium (to fit the PAMELA data), and $\Phi=750$ MV
for Carbon, Oxygen and Iron nuclei (to fit the HEAO3 data).
A higher modulation potential for HEAO3 was also found
in \cite{2011ApJ...729..106T}.

There is a ``knee'' of the all-particle spectra at PeV energies,
which indicates the existence of break or cutoff on the CR spectra
\cite{2003APh....19..193H}. Such a break or cutoff might be due to
the acceleration limit of sources, the propagation/leakage effect
from the Galaxy, or interactions with background particles
\cite{2004APh....21..241H}. Here we adopt two kinds of
cutoff/break to model the knee structure of the total spectra: a
sub-exponential cutoff case with the energy spectrum above the
injection break $R^{-\gamma_2}\exp(-R/R_c)$, and a broken
power-law case with energy spectrum above the injection break
$R^{-\gamma_2} (1+R/R_c)^{-1}$. In both cases we assume that the
cutoff/break energy is $Z$-dependent, i.e., the rigidity $R_c$ is
constant\footnote{It is reasonable to assume a constant $R_c$ for
all sources for the propagation/leakage models and the interaction
models. However, the break may suffer from a dispersion in the
acceleration limit models. We have tested that the result with a
dispersion of $R_c$ can actually be well approximated by the model
with a proper constant $R_c$.}.

The calculated energy spectra of proton, Helium, Carbon, Oxygen, Iron
and the total spectrum, together with the observational data are shown
in Fig. \ref{fig:pamela}. The parameters of the model are compiled
in Table \ref{table1}. The results show good agreement with the data
of each major species and the all particle one. Here a sub-exponential
cutoff instead of the standard exponential cutoff is required,
in order to reconcile the all-particle spectra with the proton and Helium 
data around the knee region. Such a sub-exponential cutoff may be 
originated from stochastic acceleration of particles in the turbulent 
downstream of weakly magnetized, collisionless shocks
\cite{2008ApJ...683L.163L}. For the cutoff case, the
expected all-particle spectrum is lower than the data at energies above
tens of PeV. Such a result is consistent with the requirement of a ``B
component'' of Galactic CRs \cite{2005JPhG...31R..95H}. For the break
case, the high energy data of the all-particle spectra can be well
reproduced without the ``B component'', as was also shown in the
``poly-gonato'' model \cite{2003APh....19..193H}.

\begin{table}[!htb]
\centering
\caption{Source parameters: injection spectra $\gamma_1$, $\gamma_2$ and
break rigidity $R_b$, high energy cutoff rigidity $R_c$ and solar modulation
potential $\Phi$.}
\begin{tabular}{ccccccc}
\hline \hline
  & & $\gamma_1$ & $\gamma_2$ & $R_b$ & $R_c$ & $\Phi$ \vspace{-0mm} \\
  & &            &            & (GV)  & (PV)  & (GV) \\
\hline
         & p      & $1.95\pm0.20$ & $2.52\pm0.28$ & $[5,30]$ & $0.5$ & $0.55$ \\
  cutoff & He     & $1.95\pm0.20$ & $2.50\pm0.33$ & $[5,30]$ & $0.5$ & $0.55$ \\
         & C,O,Fe & $1.95\pm0.20$ & $2.58\pm0.35$ & $[5,30]$ & $0.5$ & $0.75$ \\
  \hline
         & p      & $1.95\pm0.20$ & $2.52\pm0.25$ & $[5,30]$ & $0.5$ & $0.55$ \\
  break  & He     & $1.95\pm0.20$ & $2.50\pm0.30$ & $[5,30]$ & $0.5$ & $0.55$ \\
         & C,O,Fe & $1.95\pm0.20$ & $2.58\pm0.32$ & $[5,30]$ & $0.5$ & $0.75$ \\
  \hline
  \hline
\end{tabular}
\label{table1}
\end{table}

{\it Discussion}---In such a simple scenario of dispersion of
injection spectra of CR sources, the observed hardening of
CR spectra by ATIC, CREAM and PAMELA can be reproduced. If the CRs
are indeed originated from a population of sources instead of a single
major source (e.g., \cite{1997JPhG...23..979E,2011arXiv1101.5192G}),
such an asymptotic hardening effect due to dispersion of source
properties is inevitable. The injection parameters are similar to
those inferred from the $\gamma$-ray observations of SNRsi, which
might be evidence that SNRs are the sources of Galactic CRs below
$\sim$PV.

The injection spectra of different elements are slightly different,
which might be related to the difference in production of different
species \cite{2011ApJ...729L..13O}. We find that the difference
between various elements in the injection spectra is not as large as
that in the observed ones, possibly because the interaction strengths
of various elements are different from each other during propagation.
It should be noted that the PAMELA data actually showed a sharp
break at $\sim200$GV and a gradual softening below the break rigidity
\cite{2011Sci...332...69A}. Such detailed features cannot be simply
recovered in the present model, where the gradual hardening is expected.
The same tension also exists for the multi-component source model and the
non-linear acceleration model. We expect that future better measurements
of wide band spectra by e.g., the Alpha Magnetic Spectrometer (AMS02,
\cite{AMS02}) and the Large High Altitude Air Shower Observatory (LHAASO,
\cite{2010ChPhC..34..249C}) would help to test this model.

\begin{figure}[!htb]
\centering
\includegraphics[width=\columnwidth]{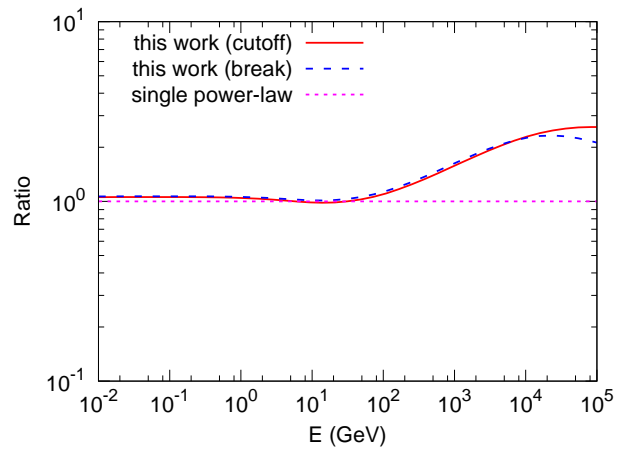}
\caption{Ratio of the hadronic component of the diffuse $\gamma$-rays
between the dispersion scenario expectation and the single power-law model.}
\label{fig:sec}
\end{figure}

There are some implications of the CR spectral hardening, for example,
the imprint on the secondary particles such as positrons
\cite{2011MNRAS.414..985L}, diffuse $\gamma$-rays, and antiprotons
\cite{2011PhRvD..83b3014D}. Based on our model CR spectra, we calculate
the predicted hadronic-origin diffuse $\gamma$-ray fluxes. The total
diffuse $\gamma$-ray emission consists of
hadronic, leptonic and the extra-galactic components, which is very
complicated. In the Galactic plane the diffuse $\gamma$-ray flux
is dominated by the hadronic component as shown in the conventional CR
propagation models \cite{2004ApJ...613..962S}. The $\gamma$-ray yield is
calculated using the parameterization of $pp$ interactions given in
\cite{2006ApJ...647..692K}. For the effect of heavy nuclei we employ a
nuclear enhancement factor $\epsilon_M=1.84$ \cite{2009APh....31..341M}.
The absolute fluxes depend on the propagation model and spatial
distribution of CRs. Here we only discuss the relative results. The
ratios of the hadronic $\gamma$-ray fluxes between
our model expectation and that of the traditional single power-law CR
spectrum are shown in Fig. \ref{fig:sec}. It is shown that the $\gamma$-ray
flux also experiences a hardening above $\sim 50$ GeV.
A similar conclusion was also derived in \cite{2011PhRvD..83b3014D}. For
positrons and antiprotons we expect similar behaviors, although
the propagation effect may change a little bit the quantitative results.
The current Fermi $\gamma$-ray data and PAMELA antiproton data can not
probe such a hardening of the secondary particles yet.

\begin{figure}[!htb]
\centering
\includegraphics[width=\columnwidth]{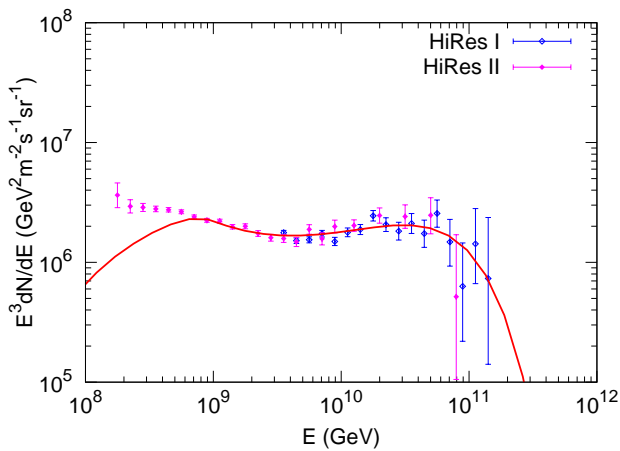}
\caption{Calculated energy spectra of UHECRs for pure protons,
compared with the HiRes data \cite{2008PhRvL.100j1101A}.}
\label{fig:uhecr}
\end{figure}

We also note that the dip-cutoff structure is very similar to the
ankle-GZK structure of ultra high energy CRs (UHECRs). Considering
the fact that UHECRs should also suffer from such a hardening effect
if they are from a population of sources, we would expect the same
mechanism to be responsible for the ankle-GZK structure of UHECRs.
Following \cite{2006PhRvD..74d3005B}, we assume the injection spectrum
of UHECRs is a broken power-law function with an exponential cutoff.
The logarithm of break energy $\log(E_b/{\rm eV})$ is assumed to be
uniformly distributed in $[17,18]$ and the cutoff energy is $E_c\sim
5\times 10^{19}$ eV. The spectral index is $2.0\pm 0.2$ below $E_b$ and
$3.6\pm 0.6$ above $E_b$. Assuming a pure proton component
\cite{2005ApJ...622..910A}, we present the expected superposed UHECR
spectrum in Fig. \ref{fig:uhecr}. We would expect a similar result
for any other single chemical species. Note that the Pierre Auger
Observatory data showed a gradual increase in mass composition
\cite{2010PhRvL.104i1101A}. Since the relative abundance of each 
chemical species is not yet well constrained
experimentally, we cannot perform a detailed modeling.
In any case, we propose that the superposition effect from
a population of UHECR sources with a dispersion in the injection
spectrum provides a new ingredient to model the UHECR spectrum.
Since the interactions between UHECRs and the background photons
are unavoidable if UHECRs are produced at cosmological distances
\cite{2006PhRvD..74d3005B}, we may in turn expect that UHECRs are
produced locally or even in the Galaxy \cite{2010PhRvL.105i1101C}
if the mechanism proposed in this work is responsible for the shape
of the UHECR spectrum. More generally, it is possible that both the
superposition and the interaction effects are in operation to give
the ankle-GZK structure of UHECRs.

\acknowledgments
This work is supported by NSF under grant AST-0908362, NASA under grants
NNX10AP53G and NNX10AD48G, and Natural Sciences Foundation of China under
grant 11075169, and the 973 project under grant 2010CB833000.


\end{document}